# Multifractal Description of Streamflow and Suspended Sediment Concentration Data from Indian River Basins

Adarsh S[1*], Drisya S Dharan[1], Nandhu AR[1], Anand Vishnu B[1], Vysakh K Mohan[1], M Wątorek[2]

[1] TKM College of Engineering Kollam, Kerala, India

[1]*Corresponding author, Adarsh S, Ph.D., Associate Professor, TKM College of Engineering Kollam, Kerala, India

adarsh_lce@yahoo.co.in; adarsh1982@tkmce.ac.in

Mob :+91-9446915388

[2]Complex Systems Theory Department, Institute of Nuclear Physics, Polish Academy of Sciences, ul. Radzikowskiego 152, 31-342 Kraków, Poland

# Multifractal Description of Streamflow and Suspended Sediment Concentration Data from Indian River Basins

Adarsh S, Drisya S Dharan, Nandhu AR, Anand Vishnu B, Vysakh K Mohan, M Wątorek

## Abstract

This study investigates the multifractality of streamflow data of 192 stations located in 13 river basins in India using the Multifractal Detrended Fluctuation Analysis (MF-DFA). The streamflow datasets of different river basins displayed multifractality and long term persistence with a mean exponent of 0.585. The streamflow records of Krishna basin displayed least persistence and that of Godavari basin displayed strongest multifractality and complexity. Subsequently, the streamflow-sediment links of five major river basins are evaluated using the novel Multifractal Cross Correlation Analysis (MFCCA) method of cross correlation studies. The results showed that the joint persistence of streamflow and total suspended sediments (TSS) is approximately the mean of the persistence of individual series. The streamflow displayed higher persistence than TSS in 60 % of the stations while in majority of stations of Godavari basin the trend was opposite. The annual cross correlation is higher than seasonal cross correlation in majority of stations but at these time scales strength of their association differs with river basin.

**Keywords**: streamflow, multifractal, sediment, persistence, correlation

## Introduction

The estimation of local fluctuations and long term dependency of hydrologic time series is a long standing problem in hydrology. Hurst exponent (Hurst 1951) is perhaps one of the most debated properties of hydro-meteorological datasets, which is mainly used to elucidate the persistence of the time series. Mandelbrot (1982) paved the way of existence of fractal geometry of geophysical fields. Over the years, a large number of methods evolved for estimation of dependency structure and fractal behavior of hydrologic time series. It includes the rescaled range analysis, double trace moments (Tessier et al. 1996), Fourier spectral analysis (Hurst et al. 1965; Pandey et al. 1998), extended self similarity principles (Dahlstedt and Jensen 2005), Wavelet Transform Modula Maxima (WTMM) (Muzy et al. 1991), arbitrary order Hilbert spectral analysis (AOHSA) (Huang et al. 2009; Adarsh et al. 2018a). Peng et al. (1994) proposed an efficient method namely Detrended Fluctuation Analysis (DFA) to perform the fractal analysis based on a detrending procedure. Kantelhardt et al. (2002) proposed the multifractal extension of DFA procedure now popularly known as multifractal DFA (MF-DFA). Multifractal is the appropriate framework for scaling fields of time series and thus can provide the natural framework for analysing and modelling various geophysical processes. For hydrological time series multifractal description can be regarded as a 'fingerprint' and it serves as an efficient nontrivial test bed for the performance of state-of-the-art precipitation-runoff models Kantelhardt et al. (2006). Therefore DFA or MF-DFA was successfully applied for characterization of various hydro-meteorological time series (Yuan et al. 2010; Yu et al. 2014; Baranowski et al. 2011; Krzyszczak et al. 2019; Adarsh et al. 2019).

Kantelhardt et al. (2003) applied the MF-DFA procedure for runoff and precipitation from different parts of globe and compared the results with WTMM method. Koscielny-Bunde et al. (2003) applied DFA, MF-DFA and wavelet analysis to discharge records from 41 hydrological stations around the globe for investigating their temporal correlations and multifractal properties. The study found that the daily runoff records were long-term correlated above some crossover time in the order of weeks, and they were characterized by a correlation function that follow a power law behaviour with exponents varying between 0.1 to 0.9. Kantelhardt et al. (2006) studied the multifractal behaviour of 99 long term daily precipitation records and 42 long term daily runoff records from different parts of the world. They found that the precipitation records generally show short term persistence while runoff records showed long term persistence with a mean exponent of 0.73. Zhang et al. (2008) applied the MF-DFA procedure to analyse the multifractal characteristics of streamflow from four gauging stations in Yangtze river in China. The study detected the non-stationarity of different time series and analysed the differences in multifractality among the records from stations at upper and lower Yangtze basin. Zhang et al. (2009) applied MF-DFA method to study the scaling behaviors of the long daily streamflow series of four hydrological stations in the mainstream of East River in China. The results indicated that streamflow series of the East River basin were characterized by anti-persistence and showed similar scaling behaviour at different shorter time scale. Further their study applied the technique to investigate the effect of water storage structures on streamflow records and found that the streamflow magnitude was mainly influenced by the precipitation magnitude while the fluctuations of the streamflow records were affected by the human interventions like construction of control structures. Labat et al. (2004) applied DFA to investigate the multifractality of streamflow series of two karstic watersheds in the southern France, suggesting that the correlation properties exist in small scales and anti-correlated properties exist in large scales. Hirpa et al. (2010) analyzed and compared the long-range correlations of river flow fluctuations from 14 stations in the Flint River Basin in the state of Georgia in the southeastern United States. The study investigated the effect of basin area on the multifractal characteristics of streamflow time series at different locations and it was found that in general, higher the basin area lower will be the degree of multifractality. Rego et al. (2013) applied the MF-DFA to analyse the multifractality of water level records of 12 principal Brazilian rivers, and the results indicated that the presence of multifractality and long-range correlations for all the stations after eliminating the climatic periodicity. Li et al. (2015) applied the MF-DFA method to the streamflow time series of four stations of Yellow river in China. They detected the crossover point at annual scale in all the time series. After removing the trend by the seasonal trend decomposition, they found that all decomposed series were characterized by the long term persistence. Also the study noted that the multifractality of streamflow series was because of the correlation properties as well as the probability density function. Tan and Gan (2017) used MF-DFA for determination of multifractal behaviour of 145 streamflow and 100 daily precipitation series of Canada. They reported that all precipitation time series showed long term persistence (LTP) at both small and large time scales, while streamflow time series generally showed LTP at large time scales. Recently, Adarsh et al. (2018b, 2019) performed the multifractal analysis of streamflow records of four stations of Brahmani river basin and one station of Kallada river basin in India.

Eventhough many studies performed multifractal characterization of streamflow employing the MF-DFA procedure worldwide, according to the author's knowledge, no comprehensive study has been reported considering streamflow data from Indian rivers and such an analysis on sediment concentration data is really scarce in literature. The specific objectives of this paper include: (i) multifractal characterization of streamflow data of different rivers in India; (ii) investigate the streamflow–suspended sediment link of five major basins in India using multifractal cross correlation analysis (MFCCA). The next section presents the theoretical details on MF-DFA and MFCCA. The details of data used in the study are presented in the section thereafter. Subsequently, results of MF-DFA analysis of streamflow and MFCCA on streamflow-total suspended sediment (TSS) links of five major basins are presented along with relevant discussions. Then the major conclusions drawn from the study are presented.

## Materials and Methods

This section presents the theoretical details on the Multifractal Detrended Fluctuation Analysis (MF-DFA) and Multifractal Detrended Cross Correlation Analysis (MFCCA) used in this study.

### Multifractal Detrended Fluctuation Analysis (MF-DFA)

The multifractal detrended fluctuation analysis (MF-DFA) is a popular tool used for the scaling characterization of non-stationary time series. The different steps involved in MF-DFA computational procedure can be described as follows:

Consider a time series $X$ ($x_1, x_2... x_N$), where $N$ is the length of the time series. The accumulated deviation of the series (known as 'profile') is calculated as:

$$X(i) = \sum_{k=1}^{i} \left[ x_k - \bar{x} \right] \quad (1)$$

where $i=1, 2,....,N$, $k=1,2...,N$, $\bar{x}$ is the mean of the series $x_k$

Divide the profile $X(i)$ into $N_s = \text{int}(N/s)$ non-overlapping segments of length, here $s$ is the segment sample size (so called scale) chosen for the analysis and int($N/s$) is the integer part of ($N/s$). As $N$ need not be a multiple of $s$ always, there is a chance of omission of small portion of the time series at the end, and to include such segments, the same procedure is repeated starting from the opposite end and a total of $2N_s$ segments are considered in the analysis

Calculate the local trend for each of the $2N_s$ segments by a least squares fit of the series as:

$$F^2(s,v) = \frac{1}{s} \sum_{i=1}^{s} \left\{ X[(v-1)s+i] - x_v(i) \right\}^2 \text{ for } v = 1,2,......,N_s \quad (2)$$

And

$$F^2(s,v) = \frac{1}{s} \sum_{i=1}^{s} \left\{ X[N-(v-N_s)s+i] - x_v(i) \right\}^2 \quad \text{for } v = N_s+1,....,2N_s \quad (3)$$

Here $x_v(i)$ is the fitting polynomial in segment $v$. Linear, quadratic, cubic etc., different types of fitting can be made and accordingly DFA procedure is named as DFA1, DFA2,.....DFAm etc.

Compute the $q^{th}$ order fluctuation function by averaging:

$$F_q(s) = \left\{ \frac{1}{2N_s} \sum_{v=1}^{2N_s} [F^2(s,v)]^{q/2} \right\}^{1/q} \tag{4}$$

Here the index variable $q$ can take any real value except zero and the zero$^{th}$ order fluctuation function is computed by following a logarithmic averaging procedure:

$$F_0(s) = \exp\left\{ \frac{1}{4N_s} \sum_{v=1}^{2Ns} \ln[F^2(s,v)] \right\} \tag{5}$$

Analyse the scaling behaviour of the fluctuation functions developing the log-log plots of $F_q(s)$ versus s for each values of $q$. If the time series is long range power law correlated, $F_q(s)$ increases as:

$F_q(s) \sim s^{h(q)}$ and $h(q)$, the slope of the plot is referred as the generalized Hurst exponent (GHE). For stationary time series, $0 < h(q = 2) < 1$, is identical to the classical Hurst exponent (Hurst 1951). For an uncorrelated series the value of Hurst exponent is 0.5. If the Hurst exponent falls between 0.5 and 1, it indicates the long term persistence (long memory process) and if it falls between 0 and 0.5, it indicates a short term persistence (short memory process). Long term persistence implies a positive autocorrelation in the time series (i.e., the effect of an observation on future observations remain significant for a long period of time). For example an extreme event would have higher probability being followed by another extreme of same character (i.e., a flood followed by another flood). The selection of scale ($s$) or segment sample size, the type of polynomial chosen etc., are some of the key issues while applying the MF-DFA method. Generally sufficient segments are chosen between the bounds (minimum and maximum) scale range. Minimum scale can be chosen in such way that it is sufficiently larger than the polynomial order chosen to prevent error in computation of local fluctuations and maximum scale below 1/10 of the sample size. Also the polynomial order can be chosen 1-3 probably sufficient to avoid overfitting problems within small segment sizes (Ihlen 2012; Oświęcimka et al. 2013).

From the GHE, several other types of scaling exponents can also be derived, which is helpful for the multifractal characterization of the time series. The $q$-order mass exponent ($\tau(q)$) and singularity exponent ($\alpha$) are derived as follows:

$$\tau(q) = qh(q) - 1 \tag{6}$$

$$\alpha = \frac{d\tau(q)}{dq} \tag{7}$$

and

$$f(\alpha) = q\alpha - \tau(q) \tag{8}$$

where $f(\alpha)$ provides the singularity spectrum. The dependency of $h(q)$ on $q$ infer multifractality of the time series and the spread of GHE plot $\Delta h(q)$ refer the strength of multifractality (Grech 2016). If the variation of GHE plot is steeper the time series is more multifractal (higher degree of multifractality) and if it is flatter the series is less multifractal (lower degree of multifractality). The base width of the singularity spectrum (spread of singularity exponent, $\Delta\alpha$) also reflects the strength of the multifractality of the time series. The shape and extent of the singularity spectrum curve contain significant information about the distribution characteristics and the singularity content of the time series. A wider singularity spectrum indicates a higher degree of multifractality and a narrow width indicates lesser degree of multifractality. For a multifractal time series the shape of singularity spectrum will be an inverted parabola whose right and left hand wings correspond to negative and positive $q$ respectively. Asymmetry Index ($A_\alpha$) is a useful parameter for multifractal analysis derived from the properties of the spectrum. It is obtained by the following relation (Drożdż et al. 2015):

$$A_\alpha = \frac{(\Delta\alpha_L - \Delta\alpha_R)}{(\Delta\alpha_L + \Delta\alpha_R)} \tag{9}$$

where $\Delta\alpha_L = \alpha_0 - \alpha_{min}$ and $\Delta\alpha_R = \alpha_{max} - \alpha_0$ are respectively, the width of left- and right-hand branches of the multifractal spectrum curve; their values describe the distribution patterns of high and low fluctuations and $\alpha_0$ is the singularity exponent for $q=0$. The value of $A_\alpha$ ranges from -1 to 1. It quantifies the deviations of the multifractal spectrum curve. $A_\alpha > 0$ suggests a left-hand deviation of the multifractal spectrum, likely to have resulted from some degree of local high fluctuations; $A_\alpha < 0$ suggests a right hand deviation with local low fluctuations, and $A_\alpha = 0$ represents a symmetrical multifractal spectrum. The difference $\Delta f(\alpha)$ between maximum and minimum values of singularity provides an estimate of the spread in changes in fractal patterns. Since $\Delta f(\alpha)$ denotes the frequency ratio of the largest to the smallest fluctuations $\Delta f(\alpha) > 0$ means that the largest fluctuations are more frequent than smallest fluctuations.

**Multifractal Cross Correlation Analysis (MFCCA)**

In order to determine the inter-relationships between different hydro-meteorological variables, different statistical approaches have been developed and simplest of which is the estimation of Pearson correlation coefficient. However, this coefficient is not robust and can be misleading if outliers are present, as in real-world data characterized by a high degree of non-linearity and non-stationarity. The Pearson correlation may display the spurious correlations in the presence of trend in non-stationary time series. Podobnik and Stanley (2008) proposed a new method, detrended cross-correlation analysis (DCCA), to investigate power-law cross correlations between two candidate non-stationarity time series in a multifractal framework. Some recent studies made detailed comparison on the person correlation and DCCA approach (Piao and Fu 2016). DCCA was extended to multifractal case and named as Multifractal Detrended Cross Correlation Analysis (MFDCCA) (Zhou 2008) and Multifractal Detrending Moving Average Cross Correlation Analysis (MFXDMA) (Jiang and Zhou 2011). Later on Oświęcimka et al. (2014) propounded a more generalized version of cross correlation analysis namely Multifractal Cross Correlation Analysis (MFCCA) which can also incorporate the sign of fluctuation function to their generalized moments. DCCA and its variants have successfully been applied to financial, biomedical and meteorological time series (Hajian and Movahed 2010; Shi 2014; Vassoler and Zebende 2012; Jiang et al. 2011; Wu et al. 2018; Dey and Mujumdar 2018).

The different steps involved in MFCCA computational procedure can be described as follows:

For two time series $x_i$ and $y_i$ ($i=1,2,…,N$); determine the profiles as two new series:

$$X(j) = \sum_{i=1}^{j} [x_i - \langle x \rangle] \qquad (10)$$

and

$$Y(j) = \sum_{i=1}^{j} [y_i - \langle y \rangle] \qquad (11)$$

where, $i = 1,2,……., N$; $\langle x \rangle$ and $\langle y \rangle$ are the mean of the two series.

Each series $x_i$ and $y_i$ are divided into $N_s$ non-overlapping segments both in progressive and retrograde directions, to avoid any omission of time series data at the beginning or end of the series. For each $2N_s$ segments, local trend of both series $x_j$ and $y_j$ are computed by fitting polynomial of appropriate order ($m$). The subtraction of the fitted polynomial from the original segment gives the covariance:

$$f_{XY}^2(\upsilon,s) = \left\{ \frac{1}{s} \sum_{k=i}^{s} \left[ (X((\upsilon-1)+k) - p_{X,\upsilon}^m(k)) \times (Y((\upsilon-1)+k) - p_{Y,\upsilon}^m(k)) \right] \right\} \quad (12)$$

Calculate detrended covariance by summing over all overlapping all segments of length $n$:

$$F^q_{XY}(s) = \frac{1}{2N_s} \sum_{\upsilon=0}^{2N_s-1} sign\left[f_{XY}^2(\upsilon,s)\right] \left|f_{XY}^2(\upsilon,s)\right|^{q/2} \quad (13)$$

$F^q_{XY}(s)$ behaves as a power-law function of $s$ (the scaling behavior), where $s$ is the segmental sample size:

$$F^q_{XY}(s) \sim s^{\lambda(q)} \quad (14)$$

The cross-correlation exponent $\lambda(q)$ similar to the generalized Hurst exponent $h(q)$ in MF-DFA and it can be obtained by observing the slope of log-log plot of $F(s)$ versus $s$ by ordinary least squares.

**Determination of Cross Correlation coefficient ($\rho_{XY}$)**

DCCA cross-correlation coefficient is defined as the ratio between the detrended covariance function $F_{XY}$ and the detrended variance functions $F_X$ and $F_Y$ (Zebende 2011; Kwapień et al. 2015)

$$\rho_{XY} = \frac{F^q_{XY}}{\sqrt{F^q_X F^q_Y}} \quad (15)$$

Theoretically the value of $\rho_{XY}$ ranges between $-1 \leq \rho_{XY} \leq 1$. If the value range between $\pm 0.666$ to $\pm 1$ cross correlation it can be considered as strong positive (or negative); $\pm 0.333$ to $\pm 0.666$ it is medium and $\pm 0$ to $\pm 0.333$ it is weak (Brito et al. 2018). The MFCCA analysis facilitate the estimation of scale dependent correlation between two candidate time series, which can provide better insight into the physical association between the variables. It is to be noted thatin this study MFCCA is retrieved for the moment order $q=2$.

## Study area and Data

In this study long term daily streamflow data of 192 stations falling in 13 river basins in India are collected from Water Resources Information System (WRIS) India (www.india-wris.nrsc.gov.in) operated by the Central Water Commission (CWC) India, which one of the most reliable database pertaining to India. The map showing different major river basins are presented in Fig. 1. The data ranging from 1969 to 2016 are considered for the study. For brevity, the maximum and minimum data lengths of the basin along with the maximum and minimum drainage area of stations of different basins, are provided in Table 1. As the total suspended sediment information is really scarce, the streamflow-sediment link is investigated in five major basins by considering the longest common period for which both the streamflow and sediment data are available.

[Insert Figure 1 Here]

[Insert Table 1 Here]

## Results and Discussions

In this study, first daily streamflow data of different stations are analysed using the MF-DFA method by selecting moment order in the range -4 to +4 and minimum scale as 10, maximum as $N/2$, where $N$ is the data length. Six different prominent multifractal properties such as Hurst exponent (H), spread of generalized Hurst exponent plot $\Delta h(q)$, spread of singularity parameter $\Delta\alpha$(called as spectral width), Asymmetry index ($A_\alpha$),$\Delta f(\alpha)$, singularity parameter for zero moment order ($\alpha_0$) etc. are evaluated. The spatial distribution of the different multifractal parameters is shown in Fig. 2. Further, the non-parametric Kernel density estimator (KDE) is used to develop the probability density function and CDF of all the six multifractal parameters andthe results are presented in Fig. 3.

[Insert Figure 2 and Figure 3 Here]

From the results it is noted that most of the streamflow series displayed long term persistence (71.3 %) with a mean value of 0.585, which is less than the universal value of 0.73 reported by Kantelhardt et al. (2006). Similarly the high multifractal width and spread ($\Delta h(q)$) are noted in the database, which shows that there is a large variation in distribution of high and low fluctuations, indicating irregular and non-homogeneous distribution. This is quite obvious because of the high intermittent character of river flows in the basins considered in the study. It is to be noted that the database considered the stations located in the southern/peninsular part of India, where in most of the rivers the streamflow is intermittent in nature and comprising of continuous zero or very low discharge values. In the northern India, abundant alluvial and perennial rivers are present, but most of them are trans-boundary in character for which the data sharing is not flexible. From Fig. 2 it is also noted that river basins Periyar, Cauvery, Pennar, Vaippar, which are near to the southern coastal regions have high degree of multifractality. The Asymmetry index value is positive for most of stations (181 stations out of 192), which indicates left hand deviations of the spectra with local high fluctuations.

From Fig. 3, it is noted that as expected the distribution of spectral width and spread (which convey the similar message on degree of multifractality) irrespective of their numerical values. The PDF of Hurst exponent shows a density concentration around 0.5-0.7, where Hurst exponent lies in this range for most of the stations (49 %). A near symmetrical distribution is noted for the value of $\Delta f(\alpha)$ and the dominant density of $\alpha_0$ is in the range of 0.8-1.2. Now, for a comparison of multifractal properties of streamflow of different basins, five major basins, namely Godavari, Krishna, Mahanadi, WFR Tadri to Kanyakumari (WFR T-K) and Cauvery are considered (for which datasets of minimum 10 stations are available). The PDFs and CDFs of different multifractal parameters are presented in Fig. 4.

[Insert Figure 4 Here]

From the PDFs and CDFs of streamflow data of river basins it is clear that the data of Krishna has least persistence (followed by Mahanadi) as compared with that of other basins. The highest degree of multifractality is noted for the streamflows of Godavari basin which is having over 400 major and minor dams and other regulation structures which control the streamflows. From Fig. 4, it is also noted that streamflows of Godavari basin has higher $\alpha_0$ as compared with other basins, which infer the complexity of the series. From the plot of $\alpha_0$ it is noted that the streamflow of Krishna and WFR T-K has almost similar complexity which possess finer structure. In the WFR T-K basin, no major flow regulation structures are present and the drainage areas of different stations are similar in magnitude (varies between 238-5755 km$^2$ from Table 1). To get an insight into the effect of drainage area and data length on the multifractality and persistence, the plots between drainage area and H drainage area and spectral width, data length vs H, data length vs spectral width are prepared and presented in Fig. 5.

[Insert Figure 5 Here]

It is evident from the Fig. 5 it is noted that that most of the Hurst exponent values are centered around 0.55-0.65 and there is no major change in the value of the Hurst Exponent with drainage area. This evidently concludes that change in drainage area has no effect on the persistence of the different series. No direct conclusions can be made from the other two plots except that area and data length independently seem to have no significant effect on the multifractality and persistence.

**MFCCA between Streamflow and Suspended Sediment**

Multifractal Cross Correlation Analysis (MFCCA) between streamflow and total suspended sediment (TSS) was performed for 5 major basins in India - Cauvery, Krishna, Godavari, Mahanadi and WFR T-K by choosing the moment order -4 to +4, maximum scale as N/2 and minimum scale is selected as more than the length of longest stretch of zero values. From the MFCCA, the individual persistence, joint persistence and cross correlation coefficient at annual and the overall correlation are determined for each case. For Cauvery basin 11 stations for which long and continuous streamflow and TSS data are available are considered for MFCCA analysis. The annual cross correlation coefficient along with Hurst exponents obtained are given in Table 2.

[Insert Table 2 Here]

Results obtained by the MFCCA analysis for streamflow and sediment data for Cauvery basin (Table 2) it is noted that the persistence of streamflow is more than that of TSS except for two stations. At all stations of Cauvery basin, the joint persistence is found to be nearly the average of individual persistence of streamflow and TSS. The joint persistence is found to be strong with a mean value of 0.733. The annual correlation is found to be more than 0.5 in five stations, but the overall correlation is found to be weak and it is less than 0.5 in all stations.The mean annual correlation is found to be 0.492 while the mean overall correlation is only 0.33. On examining the correlations it was found that, 7 out of 11 stations weak seasonal correlation (at 90 day scale) was also detected in this basin. Except for the data of Savandpur andThengumarahada stations, the annual correlation is found to be more than that of seasonal correlation. Fig. 6 shows typical plots of multifractal analysis along with the variability of cross correlation with time scale of Kudige station.

The annual and overall correlation along with Hurst exponents of datasets of different stations of Godavari basin are given in Table 3.

[Insert Table 3 Here]

[Insert Figure 6 Here]

From Table 3 it is noted that unlike for Cauvery basin, for majority of the stations in Godavari basin (i.e., 14 out of 26), the persistence of TSS is more than that of streamflow. The persistence is strong and long term for both streamflow and TSS series with a mean of 0.803 and 0.789 respectively. There exists a strong annual correlation between streamflow and TSS in this basin (mean value of 0.702). The annual correlation is greater than 0.5 in 23 cases, out of which in 17 cases the correlation is more than 0.7. The overall correlation was found to be more than 0.5 in 18 cases out of which the association is strong (>0.4) in 4 cases. For the datasets of Bishnur, Bhatpalii and Satrapur stations, both the annual and overall correlation are found to be very weak. It was also noted the seasonal correlation (at 3 month time scale) was also detectable at 16 out of 26 stations and annual correlation was found to be greater than seasonal correlation for data of all stations except Satrapur. At all stations of this basin, the joint persistence is found to be the average of persistence of streamflow and TSS. Fig. 7 shows typical plots of multifractal analysis along with the variability of cross correlation with time scale of Polavaram station in the Godavari basin.

[Insert Figure 7 here]

The annual and overall correlation between streamflow and TSS along with Hurst exponents of datasets of Krishna basin are given in Table 4.

[Insert Table 4 Here]

From Table 4 it is clear that for 14 out of 23 stations, the persistence of streamflow is more than that of TSS. In this case, the joint persistence (with a mean of 0.614) is found to be the average of the individual persistence of streamflow and TSS. Strong annual correlation (>0.7) is noted in 7 cases while it is more than 0.5 in 18 cases. In 9 cases seasonal correlation was also noted and the annual correlation is greater than that of seasonal correlation in these stations. The overall correlation was found to be weak (with a mean of 0.375) and in 5 cases the correlation is found to be more than 0.5. Fig. 8 shows typical plots of multifractal analysis of streamflow and sediment data along with the variability of Cholachguda station in Krishna basin.

[Insert Figure 8 here]

The seasonal and annual cross correlation coefficient along with Hurst exponents of datasets of Mahanadi basin is given in Table 5.

[Insert Table 5 Here]

From Table 5 it is noticed that in 81% of stations (i.e., 13 out of 16) the persistence of streamflow is more than that of TSS. Except in two cases, the seasonal correlation was detected only at Basantpur and Tikarapara station. This cross correlation coefficient is more than 0.7 at all stations except Kesinga indicating very strong positive correlation between the parameters in the basin and reasonably good overall correlation (>0.4) is noted at 14 stations. The mean value of annual correlation is found to be 0.748 while it is 0.495 for overall data. The correlation plot and multifractal plots of Basantpur station is presented in Fig. 9.

[Insert Figure 9 here]

The results of MFCCA of streamflow and TSS of WFR Tadri Kanyakumari (WFR T-K) are given in Table 6. From Table 6 it is clear that the persistence of streamflow is more that of sediment for 9 stations. The joint persistenceis nearly the mean of the individual persistence of streamflow and TSS stations of different stations. There exists reasonably good correlation at annualscalewith a mean correlation of 0.75 and the overall correlation was also more than 0.5 in 14 cases. The seasonal association was detectable at 9 station and the annual scale correlation is greater than the seasonal correlation for all the stations except for the data of Kumbidi station.

The annual cross correlation is greater than 0.5 in 18 cases out of which in 14 cases the correlation is found to be >0.7. Fig. 10 shows the multifractal plots of Ramamangalam station.

[Insert Figure 10 here]

[Insert Table 6 here]

In general, in most of the stations (57 out of 95 stations) the persistence of streamflow is greater than that of TSS. In Godavari basin, majority of the stations the persistence of TSS is more than that of streamflow. The human interventions and flow regulations might have influenced the persistence and multifractality of streamflow in this basin to a great extent. The investigation using MFCCA provides the time (scale) dependent information of the association between streamflow and TSS against the unique and traditional linear correlation between them. i.e., eventhough the overall correlation between the two are less, at specific time scale the association could be of considerable magnitude. In 45 stations, seasonal (intra-annual) association between streamflow and TSS are also noticed, among which highest number of stations (18 stations) are located in Godavari basin. This also infers the role of flow regulations in streamflow-TSS links of this basin. Eventhough streamflow-TSS association varies with temporal scales and there is no systematic pattern in this variation for the datasets of different basins. But it is noted that the strength of their association could vary significantly with time scale and their association could significantly depend on the basin and climatic (precipitation) characteristics.

## Conclusions

This study first investigated the multifractality of streamflow of 192 stations falling in 13 river basins in India using the Multifractal Detrended Fluctuation Analysis (MF-DFA). Subsequently, the Multifractal Cross Correlation Analysis (MFCCA) is employed for investigating the streamflow-sediment link in a multifractal perspective. From the results it is noted that the streamflow datasets of different river basins displayed multifractality and long term persistence with a mean exponent of 0.583. The streamflow records of Krishna basin displayed least persistence and that of Godavari displayed strongest multifractality and complexity. The streamflow-sediment links of five major river basins evaluated using MFCCA showed that the joint persistence is nearly the mean of the persistence of individual series. The streamflow displayed higher persistence than total suspended sediment in majority of the stations except that in Godavari basin. The annual cross correlation between streamflow and sediment is higher than seasonal and overall cross correlation but the strength of their association differs with river basin.

**Funding**

The authors received NO funding for performing this research work

**Conflict of Interest**

On behalf of all authors, the corresponding author states that there is no conflict of interest.

**Table 1** Details of streamflow data used for the study

| Sl No | Basin | Number of stations | Drainage Area (km$^2$) | | Data length | |
|---|---|---|---|---|---|---|
| | | | Minimum | Maximum | Minimum | Maximum |
| 1. | Krishna | 31 | 1850 | 251360 | 1095 | 18615 |
| 2. | Brahmani-Baitarani | 9 | 830 | 33955 | 4015 | 15330 |
| 3. | Sabarmati | 6 | 1421 | 19636 | 5840 | 9490 |
| 4. | Mahi | 7 | 1510 | 32510 | 3285 | 13805 |
| 5. | Mahanadi | 19 | 1100 | 124450 | 4015 | 15695 |
| 6. | Subarnarekha | 5 | 1330 | 12649 | 6205 | 14235 |
| 7. | Tapi | 5 | 8487 | 58400 | 3285 | 5110 |
| 8. | Cauvery | 31 | 258 | 66243 | 2555 | 16425 |
| 9. | WFR Tadri-Kanyakumari | 28 | 238 | 5755 | 1460 | 16425 |
| 10. | EFR Pennar-Kanyakumari | 13 | 850 | 16230 | 4015 | 16060 |
| 11. | Godavari | 23 | 2500 | 307800 | 1019 | 13111 |
| 12 | Pennar | 7 | 2486 | 37981 | 1245 | 10606 |
| 13 | WFR-Kutch-Saurashtra-Luni | 8 | 345 | 6960 | 6865 | 15111 |

Table 2 Hurst exponents of streamflow and TSS data of Cauvery basin along with the cross correlation

| Station | Hx (Streamflow) | Hy (TSS) | Scaling Exponent (Hxy) | $\rho_{XY}$ (Annual) | $\rho_{XY}$ (Overall) |
|---|---|---|---|---|---|
| Biligundulu | 0.797 | 0.669 | 0.733 | 0.404 | 0.274 |
| Kodumudi | 0.904 | 0.742 | 0.823 | 0.627 | 0.414 |
| Kollegal | 0.736 | 0.672 | 0.704 | 0.428 | 0.172 |
| Kudige | 0.745 | 0.655 | 0.700 | 0.431 | 0.249 |
| Musiri | 0.656 | 0.641 | 0.649 | 0.664 | 0.504 |
| Muthankera | 0.561 | 0.779 | 0.670 | 0.716 | 0.494 |
| Savandpur | 0.752 | 0.658 | 0.705 | 0.450 | 0.386 |
| T Narasipur | 0.781 | 0.633 | 0.707 | 0.208 | 0.090 |
| TK Halli | 0.688 | 0.673 | 0.681 | 0.566 | 0.415 |
| Tehngudi | 0.823 | 0.883 | 0.853 | 0.725 | 0.360 |
| Thengumarahada | 0.839 | 0.829 | 0.834 | 0.241 | 0.275 |

**Table 3** Hurst exponents of streamflow and TSS data of Godavari basin along with the cross correlation

| Station | Hx (Streamflow) | Hy (TSS) | Scaling Exponent(Hxy) | $\rho_{XY}$ (Annual) | $\rho_{XY}$ (Overall) |
|---|---|---|---|---|---|
| Ashti | 0.844 | 0.920 | 0.882 | 0.823 | 0.602 |
| Babli | 0.982 | 0.959 | 0.970 | 0.657 | 0.442 |
| Bamini(Balharsha) | 0.727 | 0.778 | 0.752 | 0.691 | 0.566 |
| Basar | 1.00 | 0.965 | 0.994 | 0.588 | 0.521 |
| Bhatpalli | 0.587 | 0.644 | 0.615 | 0.357 | 0.267 |
| Bishnur | 0.739 | 0.457 | 0.598 | 0.231 | 0.188 |
| Dhalegaon | 0.653 | 0.669 | 0.661 | 0.631 | 0.460 |
| G.R.Bridge | 0.767 | 0.737 | 0.752 | 0.714 | 0.561 |
| Hivra | 0.633 | 0.570 | 0.601 | 0.619 | 0.498 |
| Jagdalpur | 0.838 | 0.882 | 0.860 | 0.713 | 0.530 |
| Konta | 0.786 | 0.845 | 0.815 | 0.811 | 0.548 |
| Kumhari | 0.910 | 0.893 | 0.901 | 0.871 | 0.599 |
| Mancherial | 0.860 | 0.796 | 0.828 | 0.518 | 0.391 |
| Nandgaon | 0.724 | 0.751 | 0.737 | 0.768 | 0.614 |
| Nowrangpur | 0.850 | 0.877 | 0.864 | 0.754 | 0.662 |
| P.G. (Penganga) Bridge | 0.489 | 0.316 | 0.402 | 0.710 | 0.489 |
| Pathagudem | 0.822 | 0.894 | 0.858 | 0.886 | 0.722 |
| Pauni | 0.718 | 0.782 | 0.750 | 0.807 | 0.575 |
| Perur | 0.889 | 0.898 | 0.893 | 0.950 | 0.915 |
| Polavaram | 0.908 | 0.830 | 0.869 | 0.932 | 0.855 |
| Purna | 0.782 | 0.747 | 0.764 | 0.722 | 0.594 |
| Rajegaon | 0.998 | 1.00 | 1.000 | 0.908 | 0.549 |
| Saigaon | 0.597 | 0.562 | 0.580 | 0.769 | 0.584 |
| Satrapur | 0.957 | 0.877 | 0.917 | 0.117 | 0.277 |
| Tekra | 0.848 | 0.867 | 0.858 | 0.884 | 0.685 |
| Yelli | 0.942 | 0.965 | 0.953 | 0.809 | 0.755 |

Table 4 Hurst exponents of streamflow and TSS data of Krishna basin along with the cross correlation

| Station | Hx (Streamflow) | Hy (TSS) | Scaling Exponent(Hxy) | $\rho_{XY}$ (Annual) | $\rho_{XY}$ (Overall) |
|---|---|---|---|---|---|
| Bagalkot | 0.540 | 0.541 | 0.540 | 0.441 | 0.205 |
| Bawapuram | 0.577 | 0.505 | 0.541 | 0.644 | 0.434 |
| Byaladahalli | 0.912 | 0.870 | 0.891 | 0.813 | 0.607 |
| Cholachguda | 0.597 | 0.682 | 0.639 | 0.808 | 0.660 |
| Haralahalli | 0.751 | 0.683 | 0.717 | 0.383 | 0.296 |
| Honnali | 0.967 | 1.027 | 0.997 | 0.589 | 0.194 |
| Huvanahedgi | 0.721 | 0.650 | 0.685 | 0.241 | 0.174 |
| K Agraharam | 0.713 | 0.621 | 0.667 | 0.677 | 0.430 |
| Karaad | 0.480 | 0.449 | 0.465 | 0.659 | 0.346 |
| Keesara | 0.591 | 0.548 | 0.569 | 0.570 | 0.302 |
| Kurundwad | 0.420 | 0.487 | 0.453 | 0.938 | 0.795 |
| Malkhed | 0.655 | 0.639 | 0.647 | 0.721 | 0.210 |
| Mantralayam | 0.559 | 0.557 | 0.558 | 0.575 | 0.353 |
| Marol | 0.525 | 0.578 | 0.552 | 0.396 | 0.143 |
| Pondugala | 0.645 | 0.857 | 0.751 | 0.337 | 0.112 |
| Yadgir | 0.490 | 0.392 | 0.441 | 0.686 | 0.524 |
| Warunji | 0.655 | 0.654 | 0.654 | 0.731 | 0.493 |
| Wadanapalli | 0.675 | 0.750 | 0.713 | 0.572 | 0.336 |
| Wadakbal | 0.582 | 0.558 | 0.570 | 0.617 | 0.484 |
| Vijayawada | 0.656 | 0.590 | 0.623 | 0.702 | 0.330 |
| Takli | 0.468 | 0.365 | 0.416 | 0.504 | 0.232 |
| Shimogs | 0.557 | 0.628 | 0.592 | 0.917 | 0.686 |
| Sarati | 0.421 | 0.446 | 0.434 | 0.523 | 0.320 |

**Table 5** Hurst exponents of streamflow and TSS data of Mahanadi basin along with the cross correlation

| Station | Hx (Streamflow) | Hy (TSS) | Scaling Exponent(Hxy) | ρXY (Annual) | ρXY (Overall) |
|---|---|---|---|---|---|
| Andhiyarkore | 0.527 | 0.341 | 0.434 | 0.721 | 0.490 |
| Bamnidhi | 0.517 | 0.506 | 0.512 | 0.759 | 0.489 |
| Baronda | 0.498 | 0.416 | 0.457 | 0.757 | 0.423 |
| Basantpur | 0.691 | 0.701 | 0.696 | 0.816 | 0.552 |
| Ghatora | 1.000 | 0.991 | 1.00 | 0.782 | 0.629 |
| Jondhra | 0.537 | 0.505 | 0.521 | 0.801 | 0.513 |
| Kantamal | 0.538 | 0.415 | 0.477 | 0.726 | 0.489 |
| Kesinga | 0.99 | 1.000 | 1.00 | 0.316 | 0.364 |
| Kurubhata | 0.573 | 0.571 | 0.572 | 0.892 | 0.740 |
| Manendragarh | 0.665 | 0.777 | 0.721 | 0.779 | 0.528 |
| Rajim | 0.499 | 0.396 | 0.448 | 0.700 | 0.379 |
| Rampur | 0.483 | 0.378 | 0.430 | 0.835 | 0.484 |
| Salebhata | 0.462 | 0.386 | 0.424 | 0.763 | 0.453 |
| Simga | 0.487 | 0.400 | 0.444 | 0.720 | 0.403 |
| Sundaragarh | 0.465 | 0.387 | 0.426 | 0.833 | 0.574 |
| Tikarapara | 0.762 | 0.721 | 0.741 | 0.765 | 0.420 |

**Table 6** Hurst exponents of streamflow and TSS data of WFR-Tadri to Kanyakumari basin along with the cross correlation

| Station | Hx (Streamflow) | Hy (TSS) | Scaling Exponent (Hxy) | $\rho_{XY}$ (Annual) | $\rho_{XY}$ (Overall) |
|---|---|---|---|---|---|
| Ambarampalayam | 0.963 | 0.648 | 0.806 | 0.431 | 0.368 |
| Arangaly | 0.500 | 0.665 | 0.582 | 0.855 | 0.611 |
| Ayilam | 0.605 | 0.558 | 0.582 | 0.749 | 0.581 |
| Bantwal | 0.391 | 0.517 | 0.454 | 0.866 | 0.710 |
| Erinjipuzha | 0.722 | 0.686 | 0.704 | 0.897 | 0.669 |
| Kalampur | 0.585 | 0.636 | 0.611 | 0.786 | 0.564 |
| Kallooppara | 0.522 | 0.593 | 0.557 | 0.783 | 0.420 |
| Karathodu | 0.717 | 0.776 | 0.747 | 0.834 | 0.701 |
| Kidangoor | 0.594 | 0.803 | 0.699 | 0.657 | 0.348 |
| Kumbidi | 0.733 | 0.772 | 0.752 | 0.714 | 0.637 |
| Kuniyil | 0.584 | 0.626 | 0.605 | 0.711 | 0.589 |
| Kuttyadi | 1.000 | 0.997 | 1.00 | 0.595 | 0.415 |
| Malakkara | 0.560 | 0.645 | 0.603 | 0.678 | 0.533 |
| Neeleswaram | 1.00 | 0.916 | 0.956 | 0.858 | 0.615 |
| Pattazhy | 0.695 | 0.685 | 0.690 | 0.641 | 0.555 |
| Perumannu | 0.881 | 0.764 | 0.822 | 0.904 | 0.695 |
| Pulamanthole | 0.844 | 0.772 | 0.808 | 0.810 | 0.621 |
| Ramamangalam | 0.681 | 0.663 | 0.672 | 0.782 | 0.501 |
| Thumpamon | 0.595 | 0.674 | 0.634 | 0.730 | 0.483 |

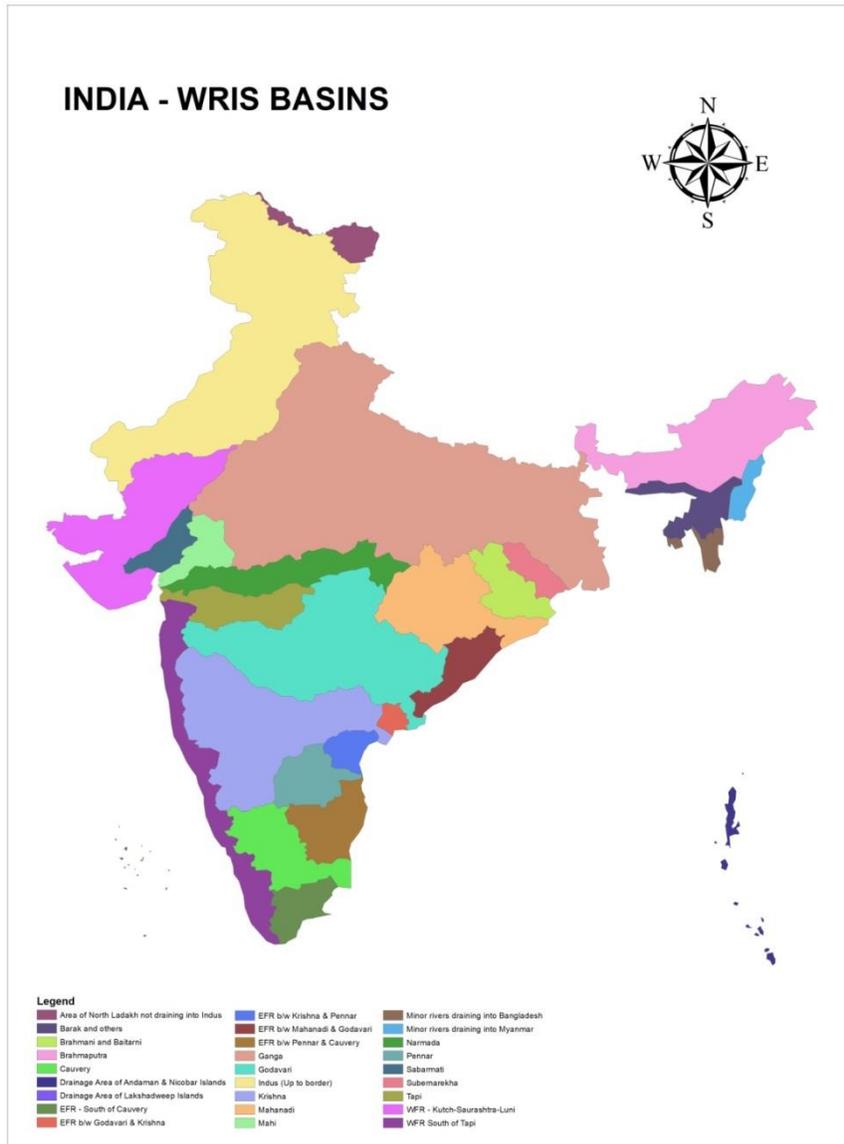

**Fig.1** Map showing river basins in India

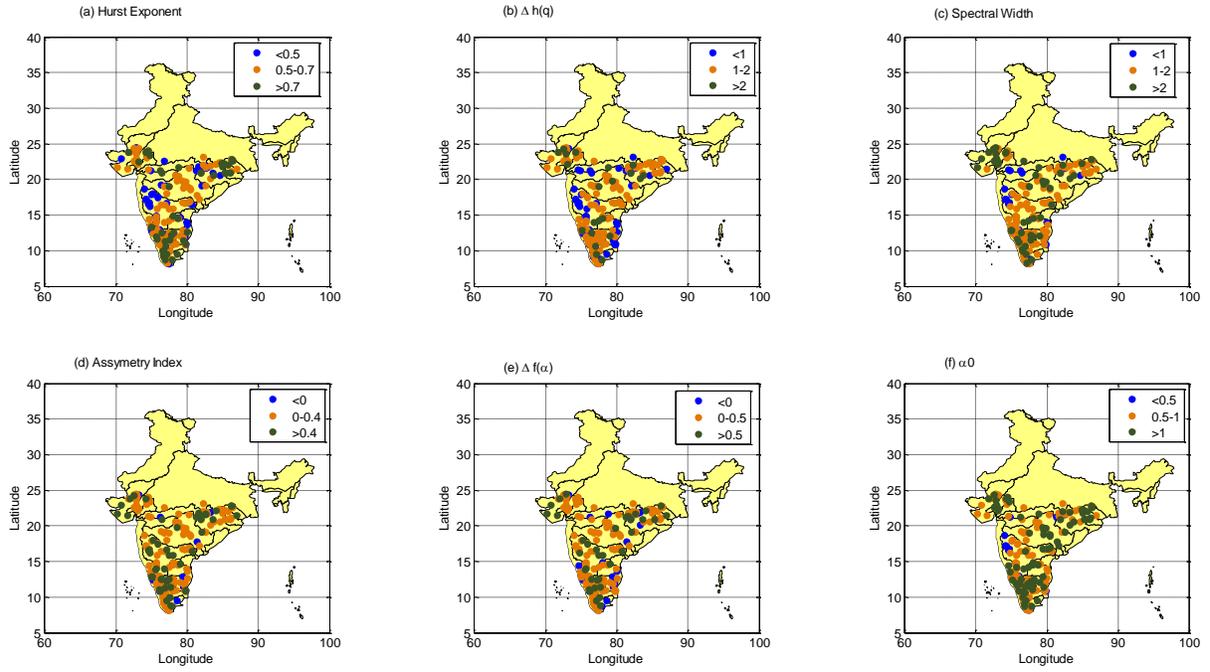

**Fig. 2** Spatial distribution of multifractal parameters of streamflow all over India (a) Hurst exponent; (b) $\Delta h(q)$; (c) spectral width; (d) Asymmetry index; (e) $\Delta f(\alpha)$; (f) $\alpha_0$

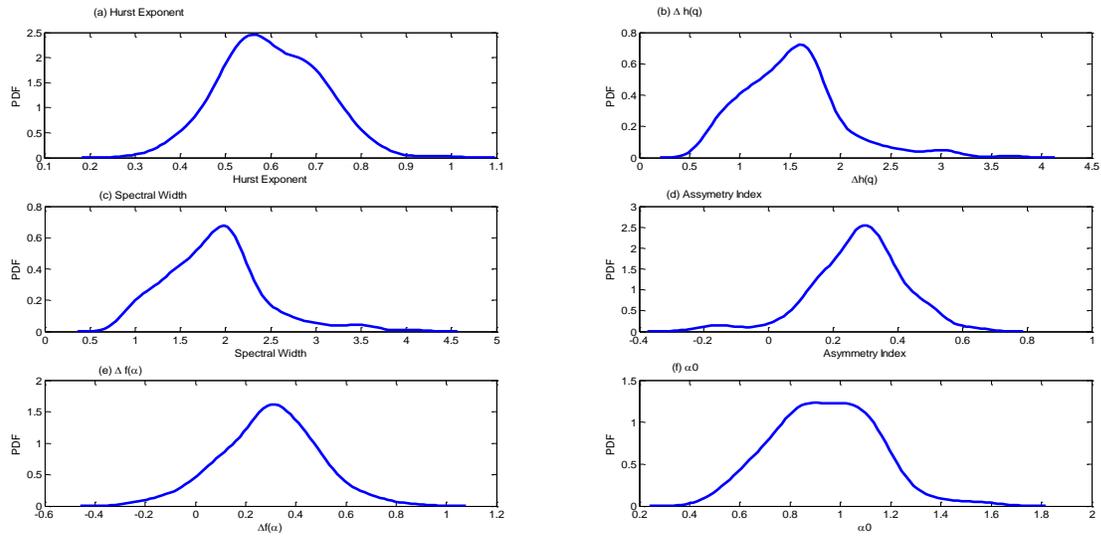

**Fig. 3.** PDF of different multifractal parameters of streamflow data

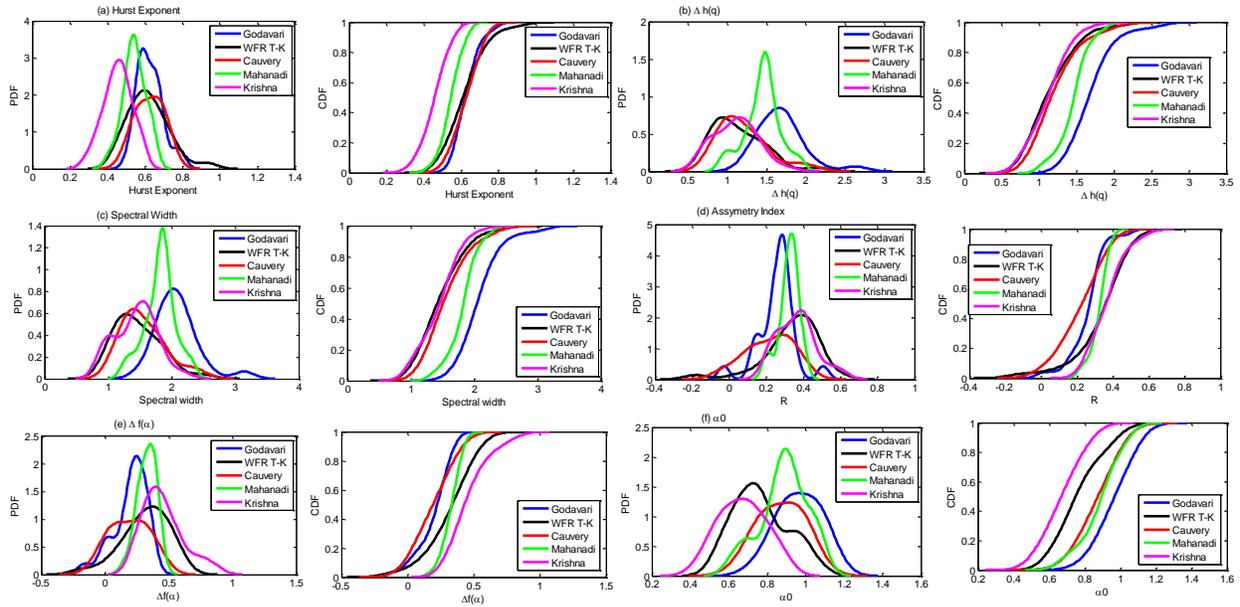

**Fig. 4** PDFs and CDFs of different multifractal parameters for basin wise analysis of streamflow datasets

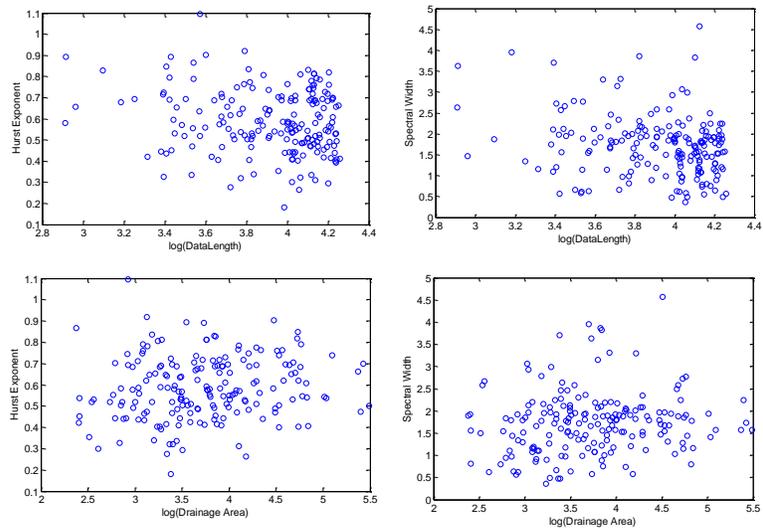

**Fig. 5** Influence of drainage area and data length on persistence and multifractality (a) Hurst Exponent vs log(Drainage Area); (b) spectral width vs log(Drainage Area); (c) Hurst Exponent vs log (Data length); (d) spectral width vs log(Data length)

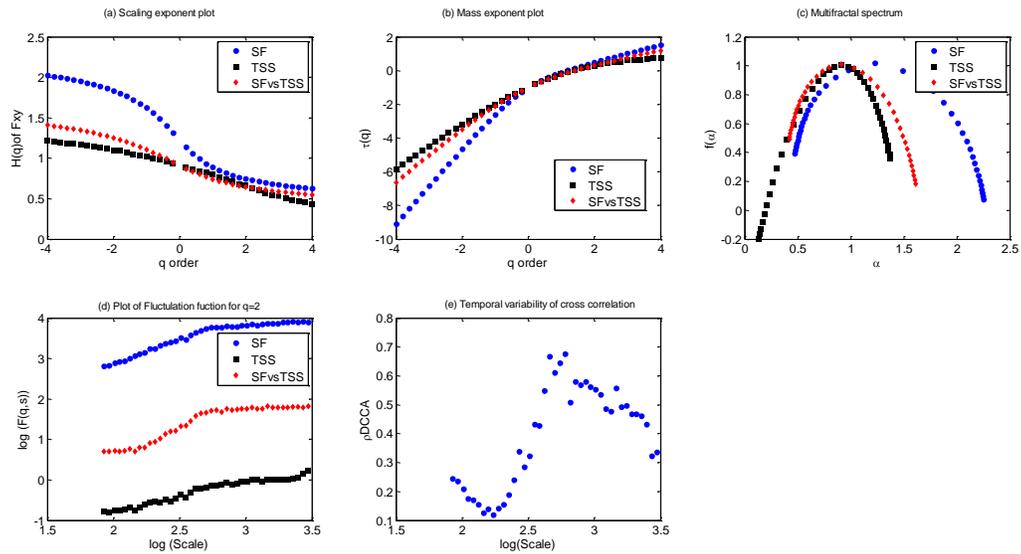

**Fig. 6** Plots of multifractal analysis of data of Kudige station along with the variability of cross correlation (a) Scaling exponent plot; (b) mass exponent plot; (c) multifractal spectrum; (d) log-log plot of fluctuation function vs scale for *q*=2; (e) temporal variability of cross correlation coefficient

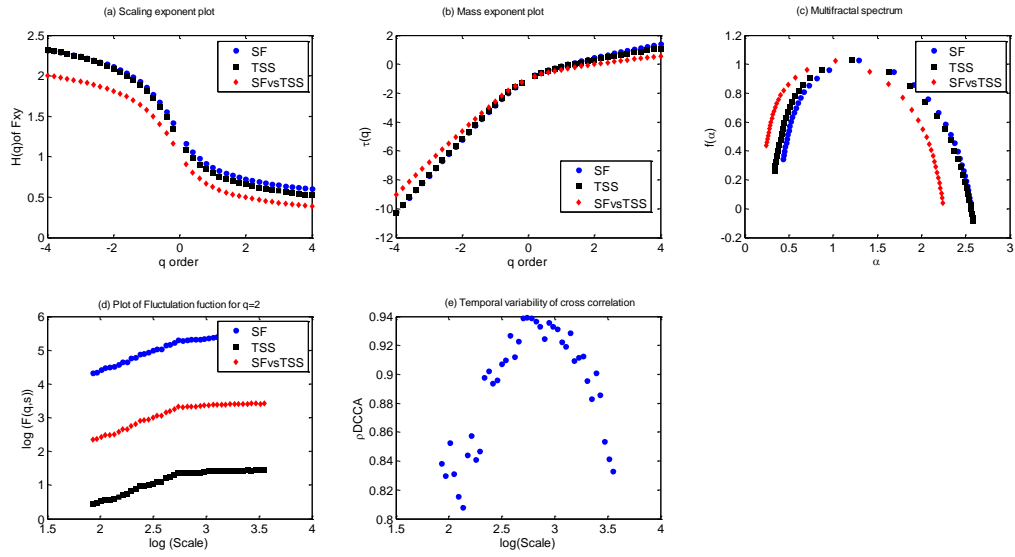

**Fig. 7** Plots of multifractal analysis of data of Polavaram station along with the variability of cross correlation (a) Scaling exponent plot; (b) mass exponent plot; (c) multifractal spectrum; (d) log-log plot of fluctuation function vs scale $q=2$; (e) temporal variability of cross correlation coefficient

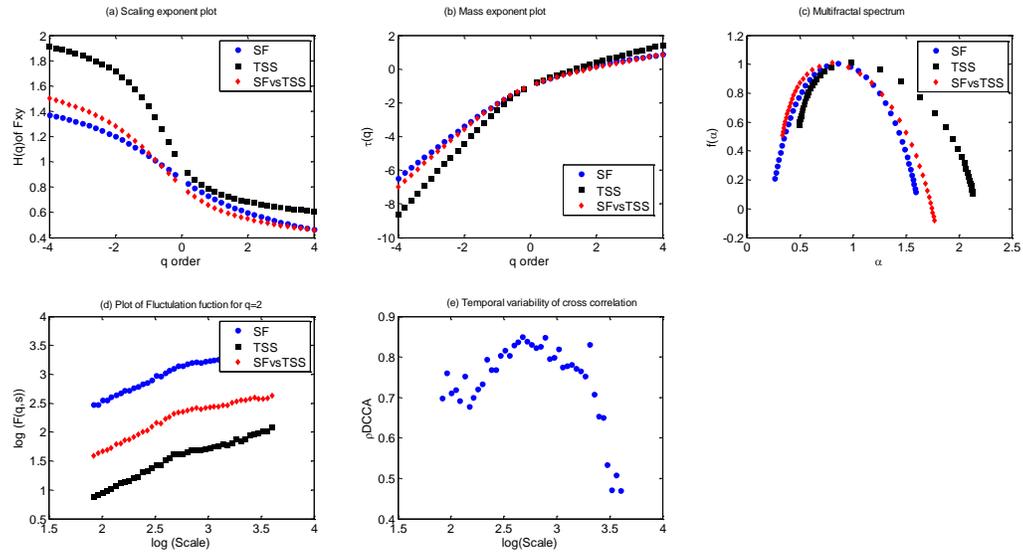

**Fig. 8** Plots of multifractal analysis of data of Cholachguda station along with the variability of cross correlation (a) Scaling exponent plot; (b) mass exponent plot; (c) multifractal spectrum (d) log-log plot of fluctuation function vs scale for $q=2$; (e) temporal variability of cross correlation coefficient

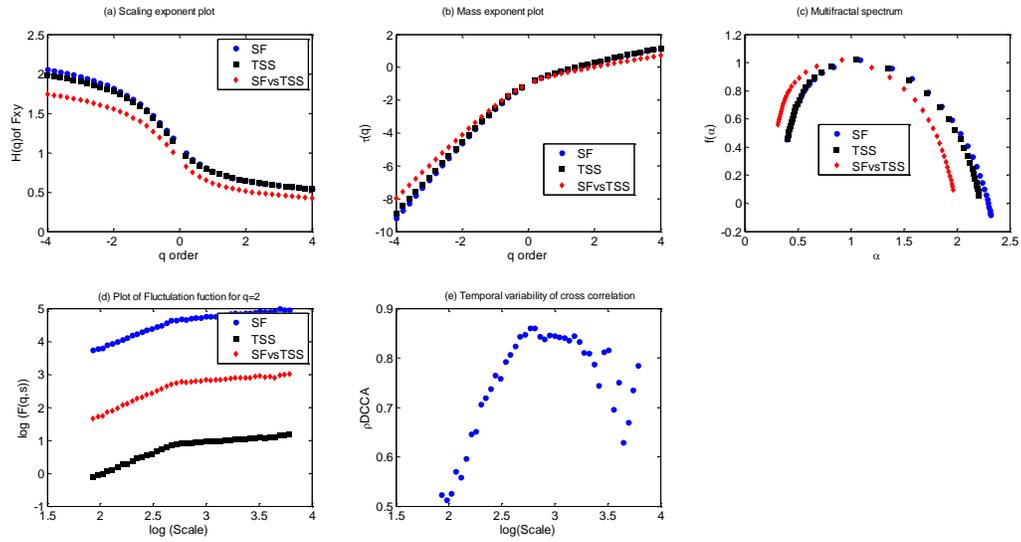

**Fig. 9** Plots of multifractal analysis of data of Basantpur station along with the variability of cross correlation (a) Scaling exponent plot; (b) mass exponent plot; (c) multifractal spectrum; (d) log-log plot of fluctuation function vs scale for $q=2$; (e) temporal variability of cross correlation coefficient

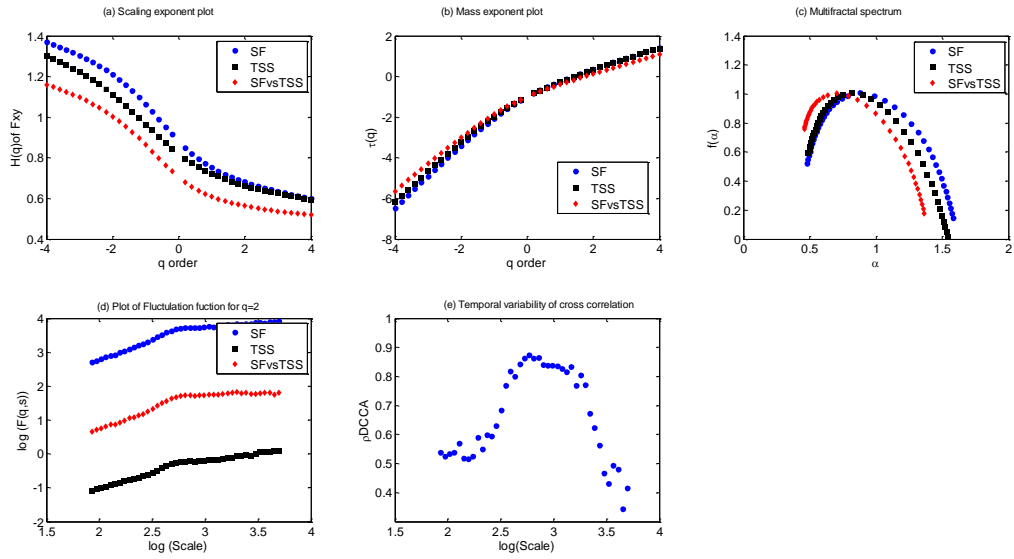

**Fig. 10** Plots of multifractal cross-correlation analysis of data of Ramamangalam station along with the variability of cross correlation (a) Scaling exponent plot; (b) mass exponent plot; (c) multifractal spectrum; (d) log-log plot of fluctuation function vs scale for $q=2$; (e) temporal variability of cross correlation coefficient